\documentclass[fleqn,12pt,twoside]{article}
\usepackage[headings]{espcrc1}
\readRCS
$Id: espcrc1.tex,v 1.2 2004/02/24 11:22:11 spepping Exp $
\usepackage{graphicx}
\usepackage[figuresright]{rotating}
\newcommand{\AmS}{{\protect\the\textfont2
  A\kern-.1667em\lower.5ex\hbox{M}\kern-.125emS}}
\newcommand{\Jpsi}{$J/\psi$}

\title{\Jpsi\ elliptic flow in relativistic heavy ion collisions}

\author{Yunpeng Liu\address[THU]{Physics Department, Tsinghua University, Beijing
100084, China},
        Nu Xu\address{Nuclear Science Division, Lawrence Berkeley
National Laboratory, Berkeley, California 94720, USA}
        and
        Pengfei Zhuang\addressmark[THU]}
\runtitle{}
\runauthor{}
\begin{document}
%
% typeset front matter
\maketitle
\begin{abstract}
The \Jpsi\ elliptic flow in high energy nuclear collisions is
calculated in a transport model. While the flow is very small at SPS
and RHIC energies, it is strongly enhanced at LHC energy due to the
dominance of the regeneration mechanism.
\end{abstract}

\section{introduction}
The $J/\psi$ production is commonly accepted as a signature of the
quark-gluon plasma formed in high energy nuclear collisions. The
$J/\psi$ elliptic flow which is sensitive to the early space-time
evolution of the colliding system carries dynamic information on the
new state of matter, if it is formed in the early stage. Since the
recently observed $J/\psi$ elliptic flow at RHIC
energy~\cite{phenix1} is with large uncertainty and can not
differentiate from various $J/\psi$ production mechanisms, it needs
more precise experimental and theoretical studies.

The charm quark number in the partonic phase increases with the
colliding nuclear energy. When the energy is high enough, the
coalescence of those uncorrelated charm quarks becomes a significant
source for $J/\psi$ production. The $J/\psi$ regeneration and its
competition with the primordial $J/\psi$ production have been
discussed in different models~\cite{thews,pbm,rapp}. To calculate
the fractions of the initially produced and regenerated $J/\psi$s
for the nuclear modification factor and transverse momentum
distribution, we set up the charmonium transport
equations~\cite{zhuang} with both gain (regeneration) and lose
(suppression) terms in the partonic phase. In this paper, we
calculate the $J/\psi$ elliptic flow in nuclear collisions at SPS,
RHIC and LHC energies in the frame of the transport model.

\section{transport model}
Taking the fact that charmonia are heavy particles, the distribution
function $f_\Psi({\bf p}_t,{\bf x}_t,\tau|{\bf b})$ where $\Psi$
stands for $J/\psi, \chi_c$ and $\psi'$ in transverse phase space at
mid-rapidity can be described by a Boltzmann equation~\cite{zhuang}
\begin{equation}
\label{transport}
\partial f_\Psi/\partial \tau +{\bf
v}_\Psi\cdot{\bf \nabla}f_\Psi = -\alpha_\Psi f_\Psi +\beta_\Psi,
\end{equation}
where ${\bf p}_t, {\bf x}_t, \tau$ and ${\bf b}$ are respectively
the charmonium transverse momentum, transverse coordinate, invariant
time and impact parameter, ${\bf v}_\Psi={\bf p}_t/E_\Psi$ is the
charmonium transverse velocity, $\alpha$ and $\beta$ are
dissociation and regeneration rate calculated from the gluon
dissociation process $\Psi+g\rightarrow c+\bar{c}$ and the reverse
process. From the large single electron elliptic flow measured at
RHIC~\cite{phenix2}, charm quarks are supposed to be kinetically
thermalized in the hot medium. The temperature, baryon chemical
potential and medium velocity appeared in the gluon and charm quark
thermal distributions are determined from the 2+1 dimensional ideal
hydrodynamics. Solving the coupled transport equation for $J/\psi$
and the hydrodynamic equations for the space-time evolution of the
partonic phase, one can obtain the $J/\psi$ distribution function
$f_{J/\psi}({\bf p}_t, {\bf x}_t, \tau|{\bf b})$ and in turn the
nuclear modification factor $R_{AA}$ and the averaged transverse
momentum square~\cite{zhuang}.

\section{$J/\psi$ elliptic flow}
\begin{figure}[!htb]
\vspace{-1cm}
 \includegraphics[height=7cm]{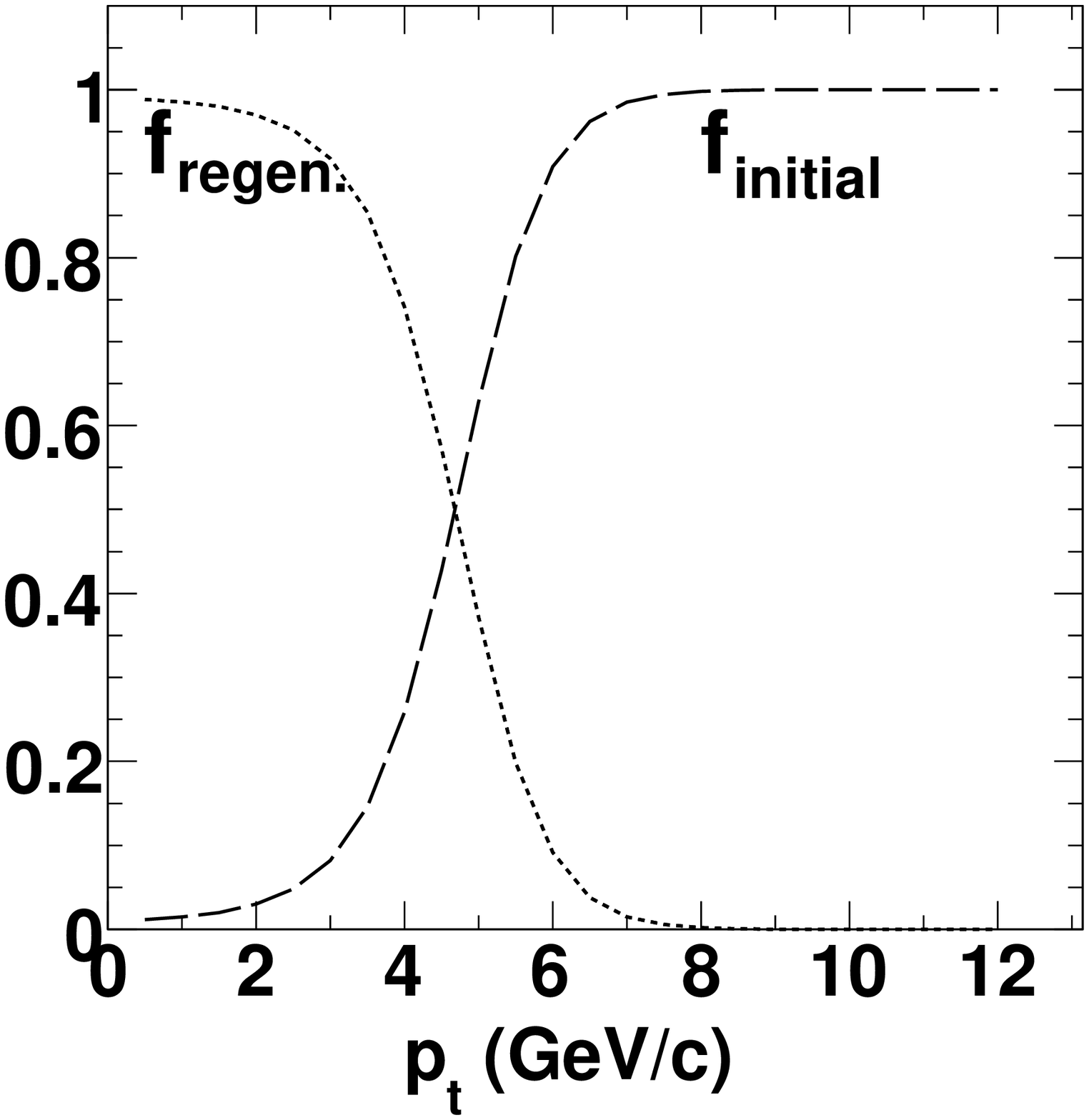}%
 \hspace{0cm}%
 \includegraphics[height=7.8cm]{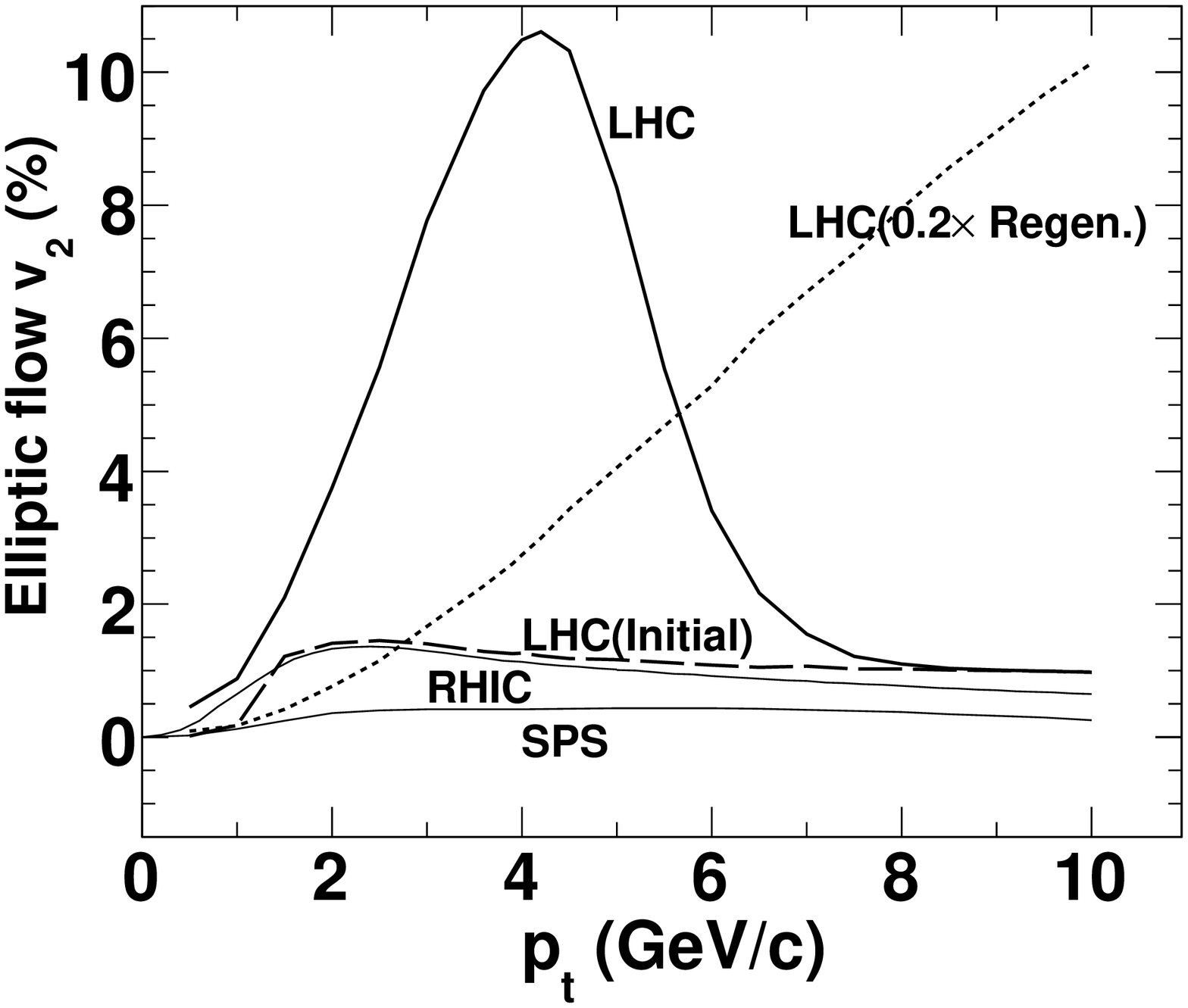}
 \vspace{-1cm}
 \caption{The $J/\psi$ production fractions (left panel) and elliptic flow (right panel)
 as functions of transverse momentum in Pb+Pb collisions at impact parameter $b=7.8$ fm
 at LHC energy. The calculations with only initial production, only regeneration
 (multiplied by a factor of 0.2) and both are
 indicated respectively by dashed, dotted and solid lines. The full elliptic flow at SPS and RHIC
 are also showed as a comparison. }
 \label{fig1}
\end{figure}

The initially produced charmonia from nucleon-nucleon hard processes
carry high momentum, and the gluon multi-scattering effect before
the two gluons fuse into a charmonium leads to a further transverse
momentum broadening~\cite{hufner}. Therefore, the high $p_t$
$J/\psi$s in the final state are dominated by the initial
production. On the other hand, the thermalized charm quarks in the
hot medium satisfy the thermal distribution, and the regenerated
charmonia are with low momentum and thus dominate the $J/\psi$
production at low $p_t$ region. The left panel of Fig.\ref{fig1}
shows the fractions of initially produced and regenerated $J/\psi$s
as functions of $p_t$ for central Pb+Pb collisions at LHC energy. It
is easy to see that the low $p_t$ region is governed by regeneration
and the high $p_t$ region is characterized by initial production.

The elliptic flow $v_2$ describes the asymmetric degree of the
particle momentum in transverse space, $v_2(p_t)=\langle
p_x^2-p_y^2\rangle/p_t^2$. At SPS energy, there are only few charm
quarks produced at the initial stage of the collisions. Therefore,
the regeneration can be safely neglected. In this case, the initial
production controls the system, and the non-thermalized $J/\psi$s
can not feel the collective flow of the thermalized medium. The only
source for the $J/\psi$ elliptic flow is the leakage effect
described by the free streaming term on the left hand side of the
transport equation (\ref{transport}): $J/\psi$s with high $p_t$ are
easier to escape the hot medium in the direction where the fireball
is thinner. Therefore, the final state $J/\psi$s are no longer
isotropic and leads to finite value of $v_2$. However, due to the
expansion of the fireball, the asymmetry in geometry decreases with
time, and the elliptic flow due to such geometry configuration is
very small in comparison with the collective effect. We show on the
right panel of Fig.\ref{fig1} the elliptic flow in heavy ion
collisions at impact parameter $b=7.8$ fm. At SPS, the elliptic flow
is extremely small due to the lack of regeneration.

The fraction of regeneration for $J/\psi$ increases with centrality
monotonously. At RHIC, the regeneration becomes important and its
contribution to the total $J/\psi$s is even around $50\%$ in central
Au+Au collisions~\cite{rapp,zhuang}. However, for semi-central
collisions where the elliptic flow reaches the maximum, the system
is still controlled by the initial production and the $J/\psi$
elliptic flow is still very weak, as showed in the right panel of
Fig.\ref{fig1}.

The situation at LHC is very different. The formed fireball is much
larger, hotter and longer lived, almost all the initially produced
$J/\psi$s at low and intermediate momentum are eaten up by the
medium, and the $J/\psi$ production with $p_t < 4$ GeV is dominated
by the regeneration, see the left panel of Fig.\ref{fig1}. In this
case the $J/\psi$s from the recombination of thermalized charm
quarks carry large elliptic flow, and the maximum value at about
$p_t \sim 4$ GeV reaches $10\%$. We have assumed charm quark
thermalization at any $p_t$, the elliptic flow from the regeneration
thus increases monotonously with $p_t$. While this is not true at
high $p_t$, it does not change the full $v_2$ remarkably, since the
high $p_t$  region is dominated by the initial production. Note that
the shape of the $J/\psi\ v_2(p_t)$ is quite different from what
observed for light quark hadrons~\cite{star} which is saturated at
high $p_t$.

In summary, the $J/\psi$ elliptic flow at LHC is expected to be much
larger than that at SPS and RHIC, due to the dominance of the
regeneration mechanism and the full thermalization of charm quarks.
\vspace{-1cm}
\section*{Acknowledgement}
The work is supported by the NSFC Grants 10735040 and 10975084, the
973-projects 2006CB921404 and 2007CB815000, and the U.S. Department
of Energy under Contract No. DE-AC03-76SF00098.

% Save this file and include it in your paper as the bibliography
% or cut and paste directly into your LaTeX


\begin{thebibliography}{20}

\bibitem{phenix1}
C.Silvestre [PHENIX Collaboration],
%``PHENIX first measurement of the J/psi elliptic flow parameter v2 in Au+Au
%collisions at sqrt(sNN) = 200 GeV,''
J. Phys. {\bf G35}, 104136 (2008).
%%CITATION = JPHGB,G35,104136;%%

\bibitem{thews}
R.Thews, M.Schroedter and J.Rafelski,
%``Enhanced J/psi production in deconfined quark matter,''
Phys. Rev. {\bf C63}, 054905 (2001).
%%CITATION = PHRVA,C63,054905;%%

\bibitem{pbm}
P. Braun-Munzinger and J. Stachel, Phys. Lett. {\bf B490},
196(2000).

\bibitem{rapp}
X. Zhao and R. Rapp,
%``Transverse Momentum Spectra of J/\psi in Heavy-Ion Collisions,''
Phys. Lett. {\bf B664}, 253 (2008).
%%CITATION = PHLTA,B664,253;%%

\bibitem{zhuang}
X. Zhu, P. Zhuang and N. Xu, Phys. Lett. {\bf B607}, 107 (2005); L.
Yan, P. Zhuang and N. Xu, Phys. Rev. Lett. {\bf 97}, 232301 (2006);
Y. Liu, Z. Qu, N. Xu and P. Zhuang, Phys. Lett. {\bf B678}, 72
(2009).

\bibitem{phenix2}
S.S. Adler et al. [PHENIX Collaboration], Phys. Rev. {\bf C72},
024901 (2005).

\bibitem{hufner}
J. Hufner, Y. Kurihara and H.J. Pirner, Phys. Lett. {\bf B215} 218
(1988).

\bibitem{star}
B.I. Abelev, et al. [STAR Collaboration], Phys. Rev. Lett. {\bf
99}, 112301(2007).
\end{thebibliography}
\end{document}